\documentclass[twocolumn,fleqn,numberedappendix,revtex4]{openjournal}
\usepackage{natbib}
\usepackage{graphicx,amsmath,amssymb,amstext}
\usepackage{amsbsy,amsfonts,amsthm,color}
\usepackage[colorlinks,linkcolor=blue,citecolor=blue,urlcolor=blue]{hyperref}
\usepackage{orcidlink}

\newcommand*    {\diff}     {\mathop{}\!\mathrm{d}}

\DeclareMathOperator{\erf}{erf}
\DeclareMathOperator{\sech}{sech}
\DeclareMathOperator{\sign}{sign}

\begin{document}
\journalinfo{The Open Journal of Astrophysics}
\submitted{submitted October 7, 2024; accepted January 8, 2025}

\setlength{\mathindent}{0pt}
\defcitealias{MiyamotoNagai1975}{MN}
\defcitealias{EvansDeZeeuw1992}{Evans-de Zeeuw}

\title{Potential-density pairs for galaxy discs with exponential or sech$^2$ vertical profile\\ \vspace*{-12mm}}

\shorttitle{Potential-density pairs for galactic discs}
\shortauthors{Dehnen \& Jafaritabar}

\author{
Walter Dehnen$^*$\,\orcidlink{0000-0001-8669-2316}}
\author{Shera Jafaritabar}

\affiliation{Astronomisches Rechen-Institut, Zentrum f{\"u}r Astronomie der Universit{\"a}t Heidelberg, M{\"o}nchhofstr.~12-14, 69120, Heidelberg, Germany}

\thanks{$^*$E-mail: \href{mailto:walter.dehnen@uni-heidelberg.de}{walter.dehnen@uni-heidelberg.de}}

\begin{abstract}
We present axially symmetric analytical potential-density pairs with surface density similar to the Miyamoto-Nagai model, but with more realistic vertical structure. Our models closely approximate an exponential, a sech$^2$, or a cored exponential vertical density profile. The latter profile has a density core of adjustable width, which provides more flexibility when modelling galaxy discs.
\end{abstract}

\keywords{methods: analytical --- galaxies: structure  --- galaxies: kinematics and dynamics \vspace*{-2mm}}

\maketitle

\section{Introduction}
\label{sec:intro}

Studies of the dynamics of the Milky Way and other disc galaxies often require models for the gravitational potential which should be realistic as well as easy to implement and compute. Unfortunately, these two goals appear mutually exclusive, since galactic discs have vertically near-exponential profiles \citep[e.g.][]{Juric2008, Dobbie2020, Mosenkov2021} for which no analytical models are known. In face of this dilemma, mainly two approaches have been used in practice: (1) numerical computation of gravitational potentials for observationally motivated density models \citep{KuijkenDubinski1995, Dehnen1998}, and (2) simple analytical disc models 
with unrealistic vertical profiles (\citealt{MiyamotoNagai1975}, hereafter \citetalias{MiyamotoNagai1975}; \citealt{EvansBowden2014}). 

The vertical density profile of the widely used \citetalias{MiyamotoNagai1975} model deviates from the exponential profile in two ways. First, at large $|z|$ it decays only like the power law $|z|^{-5}$. This deviation is presumably benign in the sense that the resulting differences between the forces at large $|z|$, and hence the ensuing orbits, are relatively minor. 

The second deviation of the \citetalias{MiyamotoNagai1975} model from a vertically exponential density profile occurs at small $|z|$, where the former exhibits a near-constant density core and declines only like $z^2$, while the latter declines linearly in $|z|$. This difference translates to the vertical Taylor expansion of the potential at small $|z|$:
\begin{align}
    \Phi =\Phi_0+\tfrac12\nu^2z^2+
    \begin{cases}
        O(|z|^3) &\text{for an exponential}, \\
        O(z^4) &\text{for a density core},
    \end{cases}
\end{align}
where $\Phi_0$ and $\nu$ denote the mid-plane value and the vertical epicycle frequency, respectively. Thus, the motion in a vertically exponential disc is much more strongly anharmonic than in a disc with density core, such as the \citetalias{MiyamotoNagai1975} model. This anharmonicity results in important dynamical effects, which are neglected when using the \citetalias{MiyamotoNagai1975} model. One effect is the presence of more orbital resonances owing to higher vertical orbital frequencies for an exponential disc. Another effect is more efficient vertical phase mixing, which in turn leads to a quick loss of vertical coherence of tidal debris from dissolving star clusters \citep{Dehnen2018} and to a faster winding of phase-spirals in the $z$-$v_z$ phase space, such as those observed in the Milky Way \citep{Antoja2018}.

A hybrid of the numerical and analytical approaches to disc modelling is to superimpose several \citetalias{MiyamotoNagai1975} models with parameters numerically determined to approximate a vertically exponential disc (\citealt{RojasNino2016} achieved $\sim10\%$ accuracy). However, this does not completely solve the discrepancies at small and large $z$, since also the combination of several \citetalias{MiyamotoNagai1975} models has an, albeit small, density core and power-law fall-off at $z\to\pm\infty$.

In this study, we introduce novel analytical potentials for galactic discs, most properties of which at $z=0$ are identical to a corresponding \citetalias{MiyamotoNagai1975} model, but which have vertical density profiles very close to exponential, sech$^2$, or a cored exponential profile. The models and their pro\-perties are derived in Sections~\ref{sec:methods} and~\ref{sec:further}, and assessed in Section~\ref{sec:assess}, while Section~\ref{sec:conclude} concludes our study.

\vspace*{2mm}
\section{Modifying the Kuzmin disc}
\label{sec:methods}
The \cite{Kuzmin1956} disc 
has gravitational potential
\begin{align}
    \Phi(R,z) = -\frac{GM}{{X}}
    \quad\text{with}\quad{X}\equiv\sqrt{R^2+Z^2},
\end{align}
where $Z=a+|z|$ with some scale length $a$, and mass density $\rho(R,z)=\Sigma(R)\delta(z)$ with surface density
\begin{align}
    \label{eq:Kuzmin:Sigma}
    \Sigma(R) = \frac{aM}{2\pi (R^2+a^2)^{3/2}}.
\end{align}
Such a razor-thin mass distribution is not very realistic, but one can obtain more realistic models by setting
\begin{align}
    \label{eq:Z}
    Z = a + \zeta(z)
\end{align}
with some function $\zeta(z)$ which in the limits $z\to\pm\infty$ approaches $|z|$, such that $\Phi\to-GM/\sqrt{R^2+z^2}$ and the parameter $M$ retains its meaning as the total mass.

The \citetalias{MiyamotoNagai1975} model is obtained from this recipe for $\zeta=\zeta_{\mathrm{M}} \equiv \sqrt{z^2+b^2}$ with scale height $b$. However, other useful but hitherto unknown modifiers $\zeta(z)$ may exist. In order to explore this possibility, we now investigate the general properties of these modified Kuzmin models and obtain conditions for the function $\zeta(z)$.

The mass density related via Poisson's equation is
\begin{align}
    \label{eq:general:rho}
    \rho(R,z) &= \frac{M}{4\pi{X}^3}\left[Z\zeta'' + \left(\frac{3Z^2}{{X}^2}-1\right) 
    (1-\zeta'^2)\right].
\end{align}
For $\zeta=|z|$, we have $\zeta'=\sign(z)$ and $\zeta''=2\delta(z)$,
such that equation~\eqref{eq:general:rho} recovers the Kuzmin model as required. In order to avoid a razor-thin component, the function $\zeta$ must be $C^2$, which for vertically symmetric discs implies $\zeta'(0)=0$. In this case, it is useful to introduce 
\begin{align}
    \label{eq:xi}
    \xi(z) \equiv [1-\zeta'^2(z)]/\zeta''(z),
\end{align}
such that the density can be written as
\begin{align}
    \label{eq:general:rho:alt}
    \rho(R,z) &= \frac{M\zeta''}{4\pi} \left[\frac{Z-\xi}{{X}^3} + \frac{3\xi Z^2}{{X}^5} 
    \right].
\end{align}
For this to be non-negative everywhere, we first require that $\zeta''\ge0$ (which with the previous conditions implies $1-\zeta'^2\ge0$ and $\xi\ge0$). In this case $\rho\ge0$ everywhere if $Z\ge\xi$. At $z=0$, this reduces to $(a+\zeta)\zeta''\ge1$, which holds for any value $a\ge0$ if $\zeta\zeta''=1$ at $z=0$, is our final condition\footnote{In view of equation~\eqref{eq:Z} and the freedom to choose $a\ge0$, we can specify $\zeta_0\equiv\zeta(0)$ without loss of generality.} for $\zeta$, and implies $\xi(0)=\zeta_0\equiv\zeta(0)$. For some models $Z<\xi$ and hence $\rho<0$ at $z\neq0$ is still possible for small $a$, as we see below.

To summarise, the function $\zeta(z)$ must satisfy the following conditions 
\begin{enumerate}
    \setlength{\topsep}{0pt}
    \setlength{\parsep}{0pt}
    \setlength{\parskip}{0pt}
    \setlength{\itemsep}{1pt}
    \item $\zeta\to|z|$ as $z\to\pm\infty$.
    \item $\zeta$ is $C^2$ with $\zeta'(0)=0$ and $\zeta''\ge0$.
    \item $\zeta''(0)=1/\zeta_0$ with $\zeta_0\equiv\zeta(0)$.
\end{enumerate}
With these conditions, the properties of modified Kuzmin models with the same values for $a$ and $\zeta_0$ are identical in the mid-plane $z=0$, regardless of their respective functions $\zeta(z)$. The potential in the mid-plane
\begin{align}
    \Phi(R,0) &=  -\frac{GM}{\sqrt{R^2+s^2}},
\end{align}
equals that of a \cite{Plummer1911} sphere with scale radius 
\begin{align}
    s \equiv Z(0) = a + \zeta_0
\end{align}
and is independent of $\zeta_0$ (at given $s$), a property inherited by its radial derivatives, e.g.\ the circular speed curve.

The \citetalias{MiyamotoNagai1975} modifier $\zeta_{\mathrm{M}}=\sqrt{z^2+b^2}$ satisfies all our conditions and gives $\xi=\zeta$ and
\citep{MiyamotoNagai1975}
\begin{align}
    \label{eq:rho:0}
    \rho_{\mathrm{M}}(R,z) &= 
    \frac{M}{4\pi}\frac{b^2}{{X}^3(z^2+b^2)} \left[\frac{a}{\sqrt{z^2+b^2}}+\frac{3Z^2}{{X}^2}\right],
\end{align}
which at fixed $R$ is near-constant for $|z|\ll b$ and declines like $|z|^{-5}$ at $|z|\to\infty$. 

To construct other useful modifiers $\zeta(z)$, we observe from equation~\eqref{eq:general:rho} that the vertical density profiles at some $R$ are close to $\zeta''(z)$ at small $|z|$. We now consider models, constructed via a recipe given in Appendix~\ref{app:recipe}, for which $\zeta''$ is either exponential or sech$^2$.

\subsection{A (nearly) exponential vertical profile}
\label{sec:exp}
A model with vertical profiles very close to exponential with scale height $h$ is generated by the modifier
\begin{subequations}
    \label{eqs:zeta:E}
\begin{align}
    \label{eq:zeta:new}
    \zeta_{\mathrm{E}}(z) = |z| + h \mathrm{e}^{-|z|/h},
\end{align}
for which
\begin{align}
    \zeta_{\mathrm{E}}'' = h^{-1}\,\mathrm{e}^{-|z|/h}
    \quad\text{and}\quad
    \xi_{\mathrm{E}} = h\big(2-\mathrm{e}^{-|z|/h}\big).
\end{align}
\end{subequations}
The resulting density 
\begin{align}
    \label{eq:rho:1}
    \rho_{\mathrm{E}}(R,z) &= 
    \frac{M}{4\pi}\frac{\mathrm{e}^{-|z|/h}}{{X}^3}\left[\frac{Z}{h}+\! \left(\frac{3Z^2}{{X}^2}-1\!\right)\big(2-\mathrm{e}^{-|z|/h}\big)\right]
\end{align}
is close to exponential at small $|z|$, but in the mid-plane is identical to the \citetalias{MiyamotoNagai1975} model for the same $s$ and $b=h$.

Since $\zeta_{\mathrm{E}}-\xi_{\mathrm{E}}<0$ at $0<|z|\lesssim1.6\,h$ with a minimum of $h(\ln2-1)$ at $|z|/h=\ln2$, the density is negative near that minimum for $h/a>1/(1-\ln2)\approx3.26$ and non-monotonic for somewhat smaller $h/a$. However, for typical applications $h\ll a$ and no such behaviour occurs.

\subsection[]{A vertical profile close to $\sech^2$}
A model with vertical profiles very close to sech$^2$ is generated by the modifier
\begin{subequations}
\begin{align}
    \zeta_{\mathrm{S}}(z) = z_0 + z_0\ln\cosh\!\frac{z}{z_0},
\end{align}
such that
\begin{align}
    \zeta_{\mathrm{S}}'' = \frac1{z_0} \sech^2\!\frac{z}{z_0}
    \quad\text{and}\quad
    \xi_{\mathrm{S}} = z_0
\end{align}
\end{subequations}
and we find from equation~\eqref{eq:general:rho:alt}
\begin{align}
    \label{eq:rho:2}
    \rho_{\mathrm{S}}(R,z) &= 
    \frac{M}{4\pi{X}^3}\sech^2\!\frac{z}{z_0}\left[\frac{a}{z_0} + \ln\cosh\!\frac{z}{z_0} + \frac{3Z^2}{{X}^2}\right]\!.
\end{align}
Again, for the same value of $s$, the mid-plane density is identical to that of the near-exponential model for $h=z_0$ or the \citetalias{MiyamotoNagai1975} model for $b=z_0$. At $|z|\gg z_0$, sech$^2(z/z_0)\sim\mathrm{e}^{-2|z|/z_0}$, such that $z_0=2h$ obtains exponential decay with scale height $h$, in which case the mid-plane density is about half of that of the exponential model.

Since $\zeta_{\mathrm{S}}-\xi_{\mathrm{S}}\ge0$, $\rho_{\mathrm{S}}\ge0$ everywhere.

\subsection{Cored exponential vertical profiles}
The sech$^2$ profile, unlike the exponential but similar to the \citetalias{MiyamotoNagai1975} disc, has a density core: a region of near-constant density at low $|z|$. Comparing
\begin{align}
    \sech^2\frac{z}{2h} = 1 - \frac{z^2}{4h^2} + O(z^4)
\end{align}
to the simple cored exponential profile
\begin{align}
    \label{eq:cored:exp}
    \mathrm{e}^{-\sqrt{z^2+w^2}/h} \propto 1 - \frac{z^2}{2hw} + O(z^4),
\end{align}
suggests $w=2h$ for the width of this core.

A cored exponential profile can be constructed as the difference of two exponentials with scale heights $h$ and $w$, respectively. Using this approach, we extend the modifier $\zeta_{\mathrm{E}}$ to include the parameter $w\in[0,h]$, generating models with cored exponential vertical profiles. For $w<h$,
\begin{subequations}
\begin{align}
    \zeta_{\mathrm{E}} &= |z| + \frac{hw(h-w) + h^3\,\mathrm{e}^{-|z|/h} - w^3\,\mathrm{e}^{-|z|/w}}{h^2-w^2},\\
    \zeta_{\mathrm{E}}' &= \mathrm{sign}(z) \left[1-\frac{h^2\,\mathrm{e}^{-|z|/h} - w^2\,\mathrm{e}^{-|z|/w}}{h^2-w^2}\right], \\
    \zeta_{\mathrm{E}}'' &= \frac{h\,\mathrm{e}^{-|z|/h} - w\,\mathrm{e}^{-|z|/w}}{h^2-w^2} \propto 1 - \frac{z^2}{2hw} + O(|z|^3),
\end{align}
\end{subequations}
which gives equations~\eqref{eqs:zeta:E} for $w=0$, while for $w=h$
\begin{subequations}
\begin{align}
    \zeta_{\mathrm{E}} &= |z| + \tfrac12\left[h + (3h+|z|)\,\mathrm{e}^{-|z|/h}\right], \\
    \zeta_{\mathrm{E}}' &= \mathrm{sign}(z)\left[1-\frac{2h+|z|}{2h} \mathrm{e}^{-|z|/h}\right], \\
    \zeta_{\mathrm{E}}'' &= \frac{h+|z|}{2h^2} \mathrm{e}^{-|z|/h}
     \propto 1 - \frac{z^2}{2h^2} + O(|z|^3).
\end{align}
\end{subequations}
For these models, $\zeta_0=h+w$, such that $s=a+h+w$. A limitation is that $w\le h$ is required (for $w>h$, $w$ and $h$ simply swap their roles).

Again, non-monotonic vertical profiles or even $\rho<0$ can occur for these models if $h\gtrsim a$.

\subsection{Modified Toomre models?}
\cite{Toomre1963} introduced a family of razor-thin discs with surface densities $\Sigma\propto(R^2+a^2)^{-k-1/2}$, which includes the \citeauthor{Kuzmin1956} disc for $k=1$. These models can be modified in exactly the same way as the \citeauthor{Kuzmin1956} disc, and \cite{MiyamotoNagai1975} gave the resulting relations for $\zeta=\zeta_{\mathrm{M}}$ and $k=1,2,3$. The density of the modified \citeauthor{Toomre1963} $k>1$ models contains a term $\propto\zeta''(\zeta-\xi)/X^3$. For the \citetalias{MiyamotoNagai1975} modifier, $\xi=\zeta$ everywhere and this term vanishes identically, but for all other modifiers, this term either causes negative densities (for $\zeta_{\mathrm{E}}$) or prevents the density from decaying faster than for the $k=1$ models at large $R$. Hence, there is little point to consider the modified $k>1$ Toomre models.\footnote{Actually, the $k>2$ \citetalias{MiyamotoNagai1975} models suffer from a similar issue: their density only decays like $R^{-5}$ at large $R$, the same as the $k=2$ model but shallower than the razor-thin $k>2$ Toomre discs.}

\section{Further analytical properties}
\label{sec:further}
The edge-on projected surface density $\Sigma=\int\!\rho\,\mathrm{d}y$ is
\begin{align}
    \label{eq:Sigma:edge:on}
    \Sigma(x,z) &= 
    \frac{M\zeta''}{2\pi(x^2+Z^2)}\left[Z + \frac{Z^2-x^2}{x^2+Z^2}\,\xi\right],
\end{align}
which in the mid-plane is again identical for models with different $\zeta(z)$ as long as they have the same values for $s$ and $\zeta_0$. The surface density $\Sigma=\int\!\rho\,\mathrm{d}z$ for the face-on projection cannot be expressed in closed form, but requires numerical treatment (see Appendix~\ref{app:surf}).

\begin{figure*}
	\includegraphics[width=18cm]{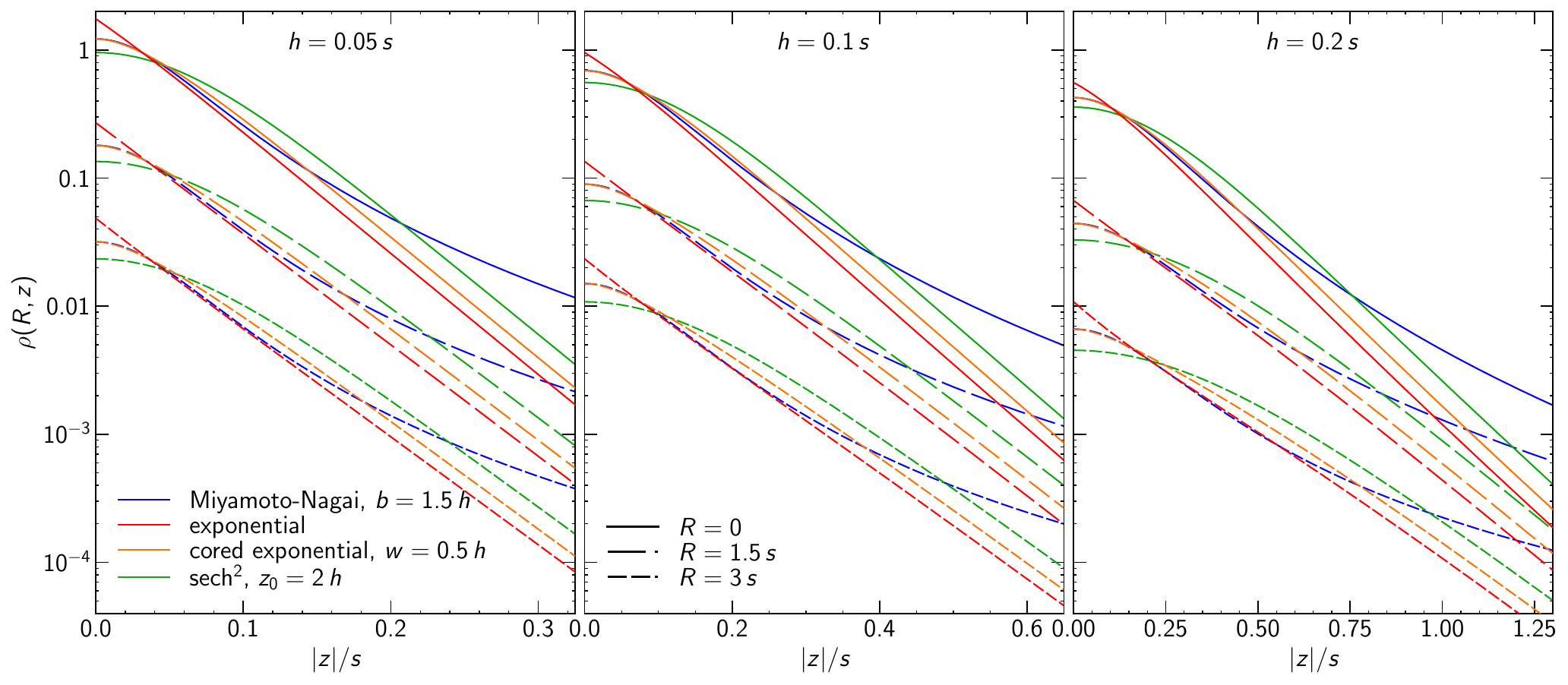}
    \vspace*{-5mm}
    \caption{
    Vertical density profiles for various modified Kuzmin models of Section~\ref{sec:methods} for different exponential scale heights $h$ and different radii $R$, as indicated. The Miyamoto-Nagai model with $b=1.5h$ approximates $\exp(-|z|/h)$ for $0.5h\lesssim|z|\lesssim3h$ and largely overlaps at $|z|\lesssim2h$ with the cored exponential model with $w=0.5h$. Note the different abscissa scales.
    }
    \label{fig:vertical_profile}
\end{figure*}

\begin{figure*}
	\includegraphics[width=18cm]{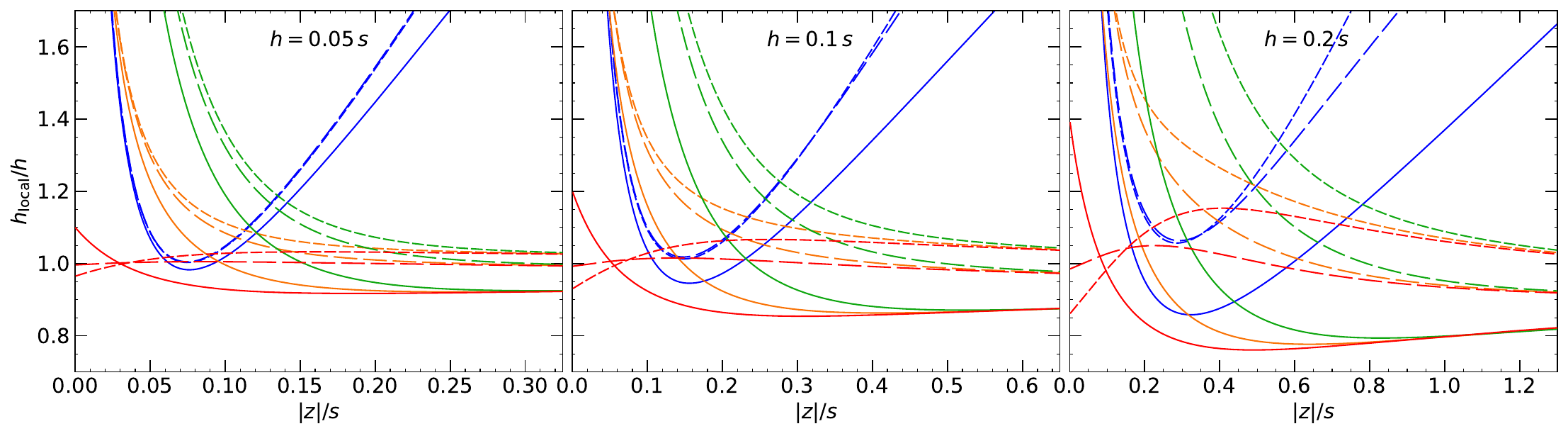}
    \vspace*{-5mm}
    \caption{
    Vertical profiles of the local exponential scale height $h_{\mathrm{local}}$ (equation~\ref{eq:h:local}) for the same models and radii as in Fig.~\ref{fig:vertical_profile} (represented with the same colours and line styles). For a perfectly exponential vertical density profile, $h_{\mathrm{local}}$ is constant. For the exponential and sech$^2$ models, $h_{\mathrm{local}}\to h$ for $z\to\pm\infty$ at fixed $R$.
    }
    \vspace*{3mm}
    \label{fig:scale-height}
\end{figure*}

The circular frequency and the radial and vertical epicycle frequencies are, respectively,
\begin{subequations}
\begin{align}
    \Omega^2(R) &= \left.\frac1R \frac{\partial\Phi}{\partial R}\right|_{z=0} &&= \frac{GM}{(R^2+s^2)^{3/2}},\\
   \kappa^2(R) &= 4\Omega^2 + R\frac{\mathrm{d}\Omega^2}{\mathrm{d}R} &&= \frac{R^2+4s^2}{R^2+s^2}\, \Omega^2(R),\\
    \nu^2(R) &=
    \left.\frac{\partial^2\Phi}{\partial z^2} \right|_{z=0} &&= 
    \frac{s}{\zeta_0} \,\Omega^2(R).
\end{align}
\end{subequations}
The first two do not depend on the vertical structure of the disc, but only on the scale radius $s$ (and total mass $M$), while the vertical epicycle frequency $\nu$ depends on the vertical structure through $\zeta_0$ and is $\sqrt{s/\zeta_0}$ times larger than $\Omega$.

\label{eqs:Jeans}
The vertical Jeans equation for axially symmetric systems reads \citep[e.g.][eq.~4.222b]{BinneyTremaine2008}
\begin{align}
    \label{eq:Jeans}
    \frac1R\frac{\partial(R\rho\overline{v_Rv_z})}{\partial R} + \frac{\partial(\rho\sigma_z^2)}{\partial z} &= - \rho \frac{\partial\Phi}{\partial z},
\end{align}
where over-lining indicates a local average and $\sigma_z^2\equiv\overline{v_z^2}$ is the vertical velocity dispersion of the population with density $\rho$. The mixed term $\overline{v_Rv_z}$ vanishes if the distribution function of that population depends only on the classical integrals energy $E$ and angular momentum $L_z$ (\citealt{NagaiMiyamoto1976}; in this case also $\sigma_R=\sigma_z$) and otherwise tends to be small, in particular for $z\ll R$. When neglecting this term, equation~\eqref{eq:Jeans} has solution
\begin{align}
    \label{eq:Jeans:int}
    \rho\sigma^2_z(R,z) &= \int_z^\infty \rho\,\frac{\partial\Phi}{\partial z}\,\mathrm{d}z.
\end{align}
In the case of self-gravitating systems ($4\pi G\rho=\nabla^2\Phi$), equation~\eqref{eq:Jeans:int} gives for the  modified Kuzmin models
\begin{align}
    \rho\sigma^2_z(R,z) &= \frac{GM^2}{8\pi} \frac{Z^2}{{X}^6}(1-\zeta'^2)
\end{align}
(see Appendix~\ref{app:sigma} for a derivation; for the \citetalias{MiyamotoNagai1975} model equivalent expressions were given by \citeauthor{NagaiMiyamoto1976} and \citealt{Ciotti1996}), such that with equation~\eqref{eq:general:rho}
\begin{align}
    \label{eq:sigma}
    \sigma_z^2 = \frac{GM}{2{X}^3} \left[\frac1Z\left(\frac1\xi - \frac1Z\right) +\frac3{{X}^2} \right]^{-1}.
\end{align}

\section{Assessing the models}
\label{sec:assess}
We now assess four examplary modified Kuzmin models: the exponential model with scale height $h$, the sech$^2$ model with $z_0=2h$, the cored exponential model with $w=0.5h$, and the \citetalias{MiyamotoNagai1975} model with $b=1.5h$. We compare these models at the same value for their respective scale radius $s$ (rather than scale length $a$), such that the mid-plane potentials (as well as circular speed and the frequencies $\Omega$ and $\kappa$) are identical between the models. The latter two models also have the same value for $\zeta_0$ such that all their properties at $z=0$ are identical. In all figures, we use units for which $G=M=s=1$.

\subsection{Vertical density profiles}
\label{sec:assess:profile}
In Fig.~\ref{fig:vertical_profile}, we plot the vertical density profiles of the models for three different values of the scale-height parameter $h$ and at three different radii $R$. We first focus on the near-exponential models (red). Their density profiles appear linear in the $\log\rho$ vs.\ $z$ presentation of Fig.~\ref{fig:vertical_profile}, implying an exponential decay. However, the slopes at different radii $R$ are not exactly the same but become shallower for larger $R$ (corresponding to a flaring disc), in particular for larger $h$. Moreover, as already discussed, the profiles are not exactly exponential, but display some deviation, most strongly at $|z|<h$.

In order to assess these deviations, we plot in Fig.~\ref{fig:scale-height} the local exponential scale height 
\begin{align}
    \label{eq:h:local}
    h_{\mathrm{local}} = \left|\frac{\partial\ln\rho}{\partial z}\right|^{-1}
\end{align}
as function of $z$. For a perfectly exponential profile, $h_{\mathrm{local}}$ would be constant, namely $h_{\mathrm{local}}=h$. The (nearly) exponential model (red) deviates slightly from these ideals, the stronger the larger $h$. When modelling the thin stellar disc of, say, the Milky Way, $h\simeq0.05s$ is a reasonable choice\footnote{The surface density of these models is close to that of the \citeauthor{Kuzmin1956} disc, whose local exponential scale length $h_R$ (computed equivalently to $h_{\mathrm{local}}$) has the minimum $2s/3$ at $R=s$. Hence, one may assign a scale radius for a modified Kuzmin model as $s=1.5h_R$. Thus, $h=0.05s$ corresponds to $h_R/h_z=3.2\mathrm{kpc}/240\mathrm{pc}$.}, for which these deviations are quite small.

Next, we consider the cored exponential (orange in Fig.~\ref{fig:vertical_profile}) and sech$^2$ (green) models. At large $z$, their vertical profiles, plotted in Fig.~\ref{fig:vertical_profile}, have the same slope and hence scale height as the exponential model, while $h_{\mathrm{local}}\to\infty$ at $z\to0$ as a consequence of the density core.

Finally, we also plot in Figs.~\ref{fig:vertical_profile} and \ref{fig:scale-height} the \citetalias{MiyamotoNagai1975} models for $b=1.5h$ (blue), which for $|z|\lesssim 3h$ are very similar to the cored exponential models with $w=0.5h$, but deviate increasingly at larger $|z|$,
where they only decay as $|z|^{-5}$.

\subsection{Projected density}
\label{sec:assess:surface:density}
In the limit $h\to0$ (or $b\to0$) the modified Kuzmin models approach the \citeauthor{Kuzmin1956} disc and hence their face-on projected surface density $\Sigma(R)=\int\rho\,\mathrm{d}z$ converges towards equation~\eqref{eq:Kuzmin:Sigma}. Therefore, we expect $\Sigma(R)$ to deviate only slightly from that of the \citeauthor{Kuzmin1956} disc as long as $h\ll s$.

\begin{figure*}
	\includegraphics[width=18cm]{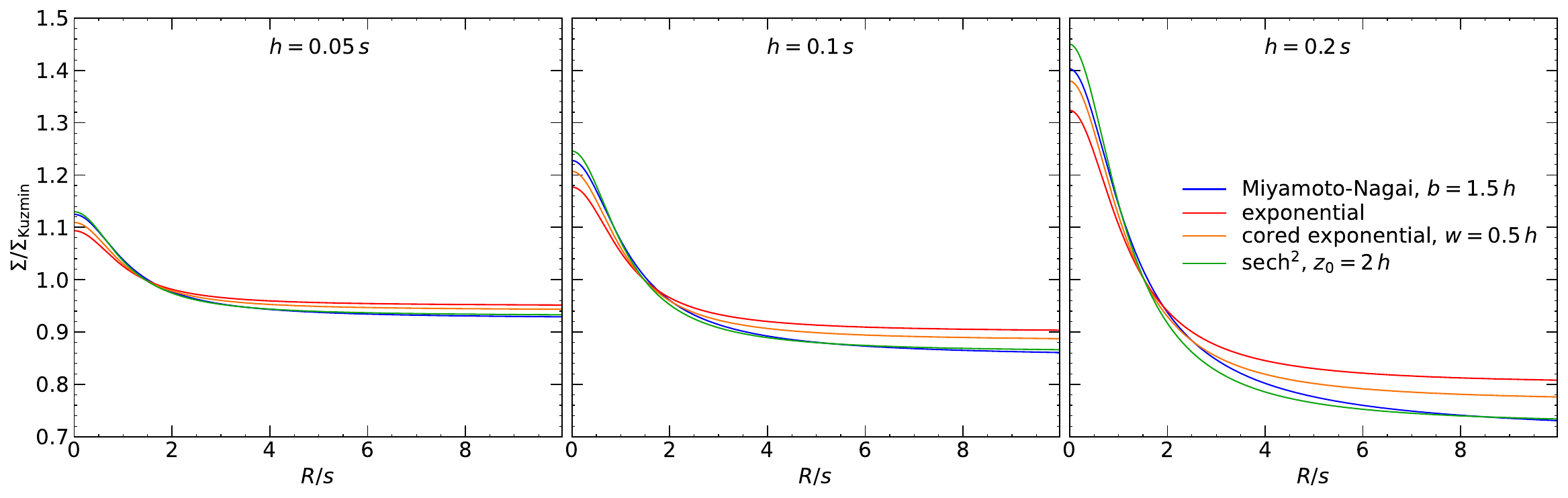}
    \vspace*{-5mm}
    \caption{
    Radial profiles of the ratio of the (face-on projected) surface density $\Sigma(R)$ to that of the \citeauthor{Kuzmin1956} disc (equation~\ref{eq:Kuzmin:Sigma} with $a=s$) for the same models as in Fig.~\ref{fig:vertical_profile}. 
    }
    \vspace*{4mm}
    \label{fig:surface_density}
\end{figure*}

\begin{figure}
	\includegraphics[width=\columnwidth]{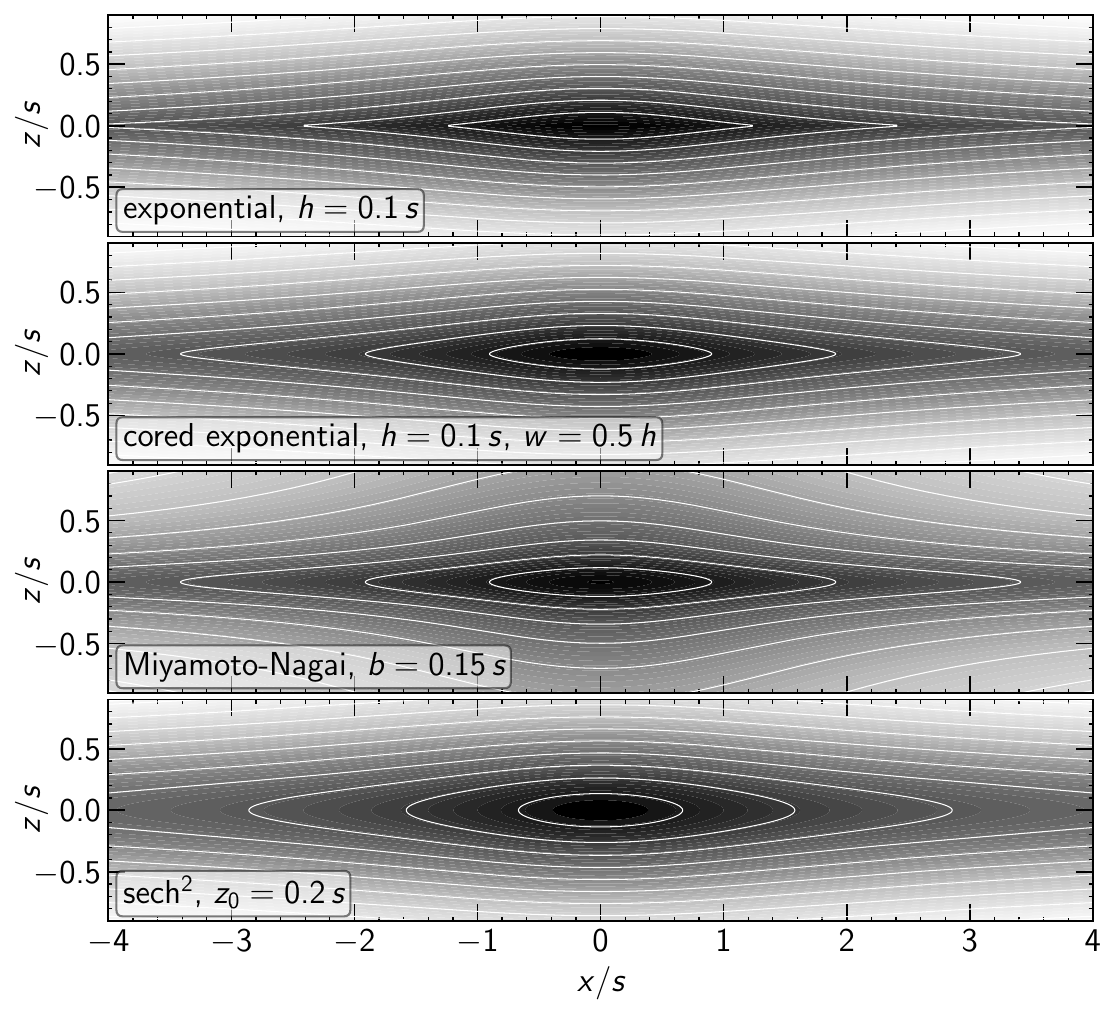}
    \vspace*{-5mm}
    \caption{
    Contours (logarithmically spaced) of the edge-on projected density for the same models as shown in Fig.~\ref{fig:vertical_profile}. Contour levels and grey-scale map are the same for all models. Note the behaviour of the contours near $z=0$ and towards large $|z|$.}
    \label{fig:projected_density}
\end{figure}

In Fig.~\ref{fig:surface_density}, we plot the ratio of $\Sigma(R)$ (computed numerically as detailed in Appendix~\ref{app:surf}) to the surface density~\eqref{eq:Kuzmin:Sigma} of the \citeauthor{Kuzmin1956} disc with the same scale radius $s$. We find indeed that these ratios are close to unity for small $h$, but deviate from this with increasing $h$. In each case, the model with surface density closest to the \citeauthor{Kuzmin1956} disc is the exponential model, while cored models deviate more, the stronger the more cored they are. However, at the same value for $h$ (so that the models have comparable exponential scale heights) the surface density between the models are very similar in the sense that they differ much more from the \citeauthor{Kuzmin1956} disc than from each other.

\begin{figure}
	\includegraphics[width=83mm]{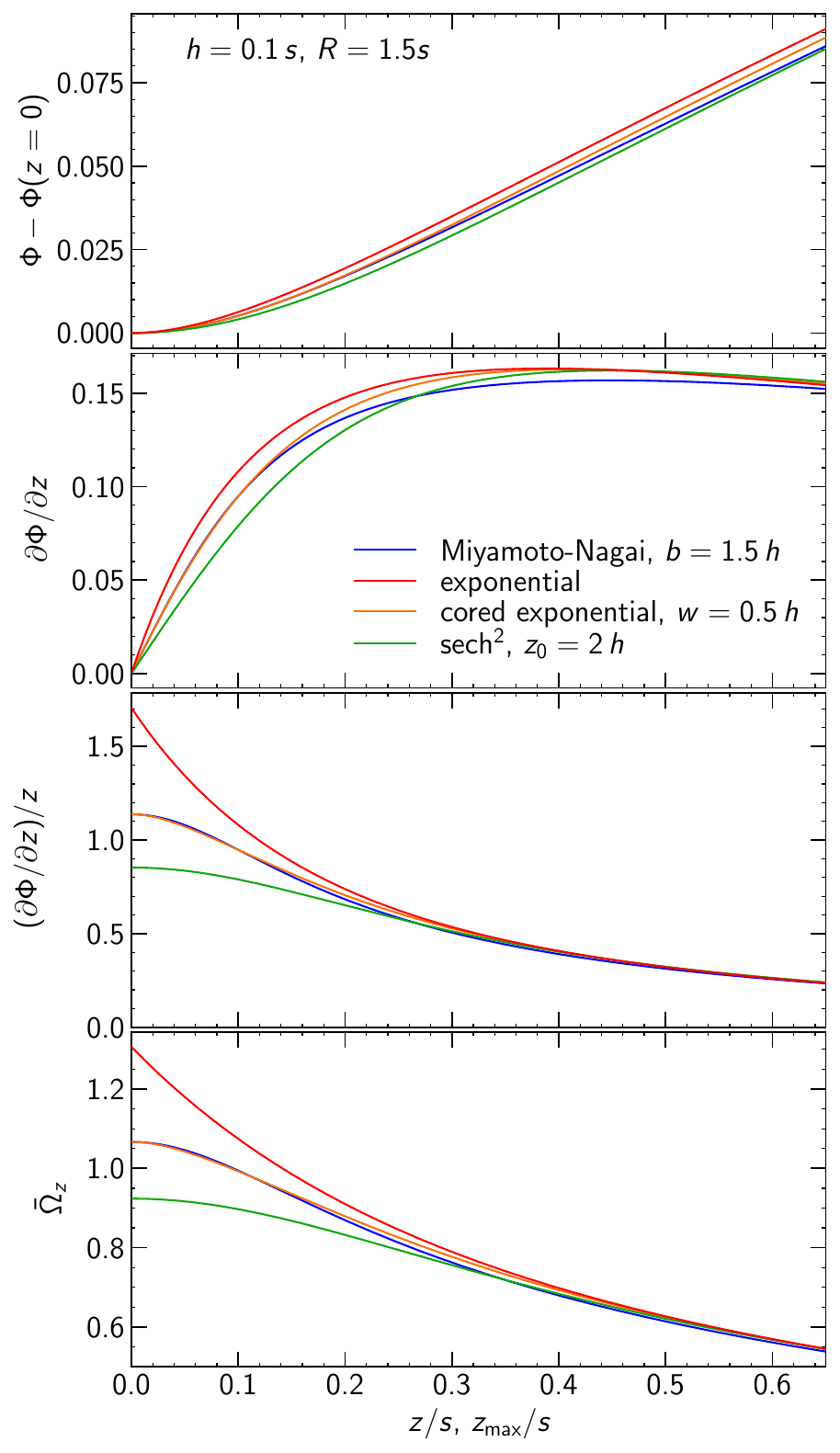}
    \vspace*{-2mm}
    \caption{
    Profiles (from top to bottom) of the gravitational potential, the (negative) acceleration, and acceleration over height $z$ as function of height, as well as that of the approximation~\eqref{eq:OmegaZ:1D} of the vertical orbital frequency as function of the maximal orbital height $z_{\max}$. Models are the same as in other figures with $h=0.1s$ and at $R=1.5s$ (e.g.~the middle panels of Figs.~\ref{fig:vertical_profile} and~\ref{fig:scale-height}, which also have the same $x$-axis scale). Units are such that $GM=1=s$.}
    \vspace*{2mm}
    \label{fig:vertical_force}
\end{figure}

In Fig.~\ref{fig:projected_density}, we plot the contours of the edge-on projected densities (equation~\ref{eq:Sigma:edge:on}) for the choice $h=0.1s$. The models differ in two aspects. First, the shape of the contours at $z=0$ is pointed for the exponential profile but rounded for all the cored profiles (this difference may in practice be hard to observe owing to internal dust obscuration). Second, the projected density of the \citetalias{MiyamotoNagai1975} disc declines much more slowly at large $|z|$.

\subsection{Vertical gravity and orbital frequency}
\label{sec:assess:force}
As emphasised in the introduction, an important difference between the \citetalias{MiyamotoNagai1975} disc and a vertically exponential disc is the behaviour of gravity at small heights $z$. To demonstrate this, we plot in the upper three panels of Fig.~\ref{fig:vertical_force} the potential, acceleration, and acceleration over height as function of height for the same four models. While the potentials look rather similar, the vertical force of the exponential model at small $|z|$ is clearly stronger than for the cored models, an immediate consequence of the larger amount of mass at near $z=0$. This difference to the cored models is more obvious in the plots of the ratio $(\partial\Phi/\partial z)/z$ of acceleration to height, which for exactly harmonic potentials is constant. The cored models indeed possess a small region of near-constant $(\partial\Phi/\partial z)/z$ around the mid-plane.

In the bottom panel of Fig.~\ref{fig:vertical_force}, we plot the approximation
\begin{align}
    \label{eq:OmegaZ:1D}
    \bar{\Omega}_z(R,z_{\max}) = \frac\pi2 \bigg/\!\!\!\int_0^{z_{\max}}\!\!\frac{\mathrm{d}z}{\sqrt{2[\Phi(R,z_{\max})-\Phi(R,z)]}}
\end{align}
for the vertical orbital frequency $\Omega_z$ as function of maximum orbital height $z_{\max}$\footnote{For small $z_{\max}$, $\bar{\Omega}_z$ computed at an average radius $\bar{R}$ is a good approximation for the actual orbital frequency $\Omega_z$.}. Its profiles are reminiscent of those for $(\partial\Phi/\partial z)/z$ (which at $z\to0$ converges to $\nu^2=\bar{\Omega}_z^2(z_{\max}=0)$), in particular, for the cored models $\bar{\Omega}_z$ is near-constant at small $z_{\max}$, while the exponential disc shows no such behaviour. The degree of orbital phase mixing is determined by the gradient $\mathrm{d}\Omega_z/\mathrm{d}z_{\max}$, which for the exponential disc is largest at small $z$, where it vanishes for the cored models. This implies that the behaviour of vertical phase mixing, which drives the evolution of the $z$-$v_z$ phase-spiral in the Milky Way \citep{BinneySchoenrich2018}, is fundamentally different between these two models.

We also see from the bottom panel of Fig.~\ref{fig:vertical_force} that at small $z_{\max}$ an exponential profile reaches larger frequencies $\Omega_z$ than a cored profile with the same exponential scale height. This has important implications for the existence of the $\Omega_z$:$\,\Omega_r=2$\,:\,1 and 3\,:\,2 orbital resonances, which, depending on the width of the core, may not occur in cored models or only over a smaller radial range than for the purely exponential models.

\begin{figure}
	\includegraphics[width=83mm]{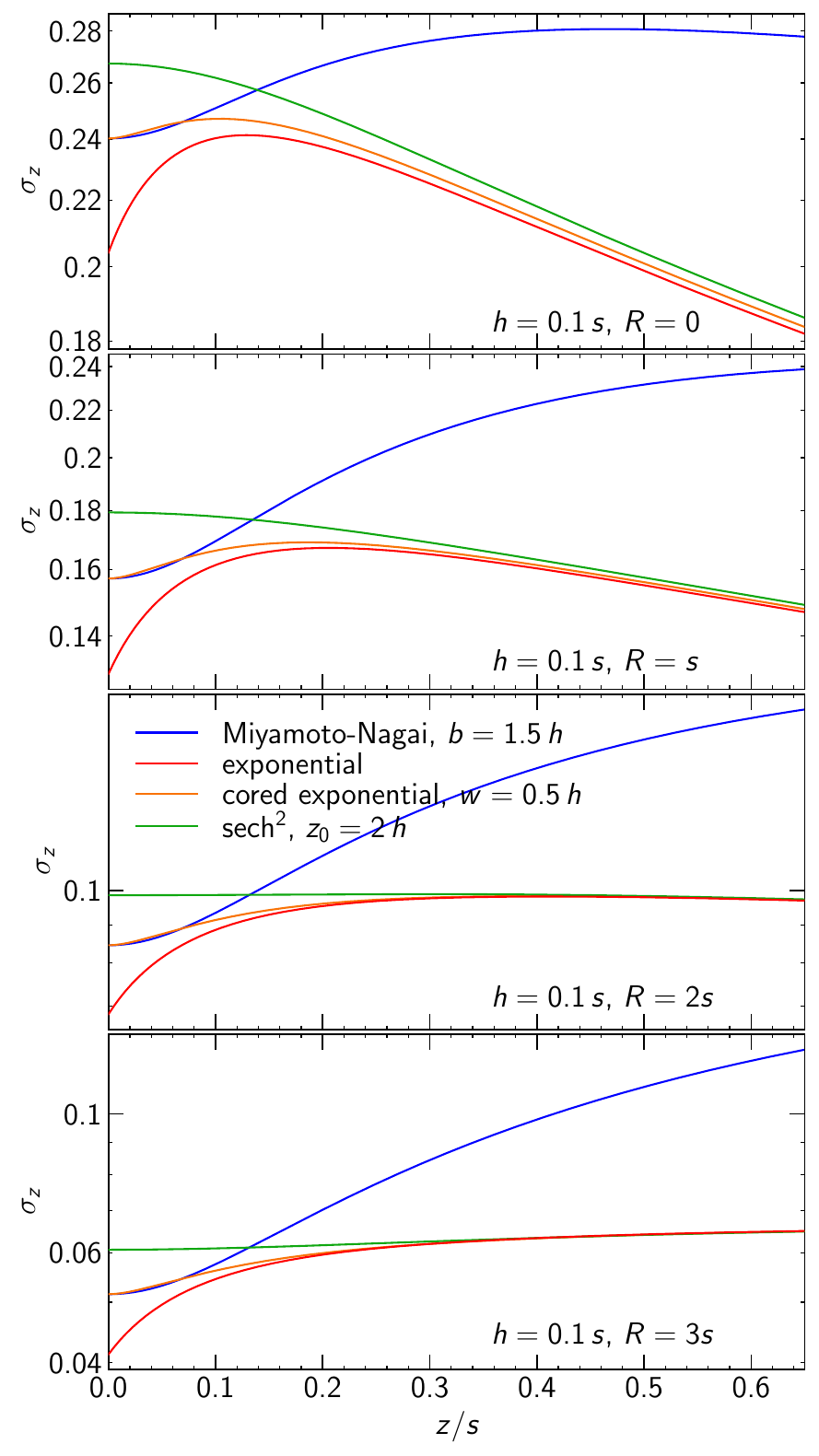}
    \caption{
    Profiles of the vertical velocity dispersion $\sigma_z$ for self-gravitating modified Kuzmin models at different radii.
    Units are such that $GM=1=s$.}
    \vspace*{2mm}
    \label{fig:velocity_dispersion}
\end{figure}

\subsection{Velocity dispersion}
For small heights $|z|$, i.e. for most disc stars, the vertical Jeans equation~\eqref{eq:Jeans} is well approximated by neglecting (i) the mixed-derivative term, since it is generally much smaller than the dominant terms, and (ii) gravity from spheroidal components, since their vertical force is much smaller than that from the disc. Hence, equation~\eqref{eq:sigma}, which is obtained under these approximations, gives a good description of the vertical velocity dispersion $\sigma_z$ of a stellar disc following a modified Kuzmin model.

In Fig.~\ref{fig:velocity_dispersion}, we plot the profiles $\sigma_z(z)$ for our four comparison models for $h=0.1s$ at different radii $R$. Consider first the sech$^2$ model (green in Fig.~\ref{fig:velocity_dispersion}), which according to simple one-dimensional theory should produce a constant $\sigma^2_z=8\pi G h^2 \rho(R,0)$ \citep[][see also problem 4.21 of \citealt{BinneyTremaine2008}]{Spitzer1942}. At small $|z|$, this model indeed has near-constant $\sigma_z$ (though not at the value expected from one-dimensional theory) but typically does not remain constant with height. These deviations are mainly owed to the departure from the assumption (used in the simple theory) of a reduced, one-dimensional Poisson equation, $4\pi G\rho=\partial^2\Phi/\partial z^2$.

All the models with an exponential vertical profile asymptote to very similar $\sigma_z$ profiles at large $|z|$, but differ at small $|z|$. The purely exponential model shows a sharp minimum for $\sigma_z$, which is reminiscent of the profile observed in the Solar neighbourhood \citep{Fuchs2009}, while the cored exponential has a less pronounced and smooth minimum.

The \citetalias{MiyamotoNagai1975} model has a completely different $\sigma_z$ profile than the exponentially declining models and reaches much larger values with maxima not captured in Fig.~\ref{fig:velocity_dispersion} (except for $R=0$). This deviation is hardly due to the different gravity, but mainly owed to the shallower density profile: $\sigma_z$ is inflated by the high speeds of stars visiting from $z_{\max} \gg |z|$, but hardly present in exponentially decaying models.

\vspace*{3mm}
\section{Discussion and conclusion}
\label{sec:conclude}
The main results of this study are the analytic mass models for thin discs with exponential or sech$^2$ vertical profile. These are obtained by modifying the razor-thin \cite{Kuzmin1956} disc, very similar to how the \citeauthor{MiyamotoNagai1975} (\citeyear{MiyamotoNagai1975}; \citetalias{MiyamotoNagai1975}) model is constructed. In fact, many properties of the new models closely follow those of the \citetalias{MiyamotoNagai1975} model (with appropriately chosen parameters), so that the main difference is the vertical structure. Our approach can be used to construct modified Kuzmin models with arbitrary vertical profiles and in Appendix~\ref{app:recipe} we give the relations for a vertically near-Gaussian model.

The central density of an exponential vertical profile is significantly higher than for the \citetalias{MiyamotoNagai1975} or the sech$^2$ models (at the same surface density and exponential scale height). This higher central density results in a stronger vertical force at small $|z|$, and therefore also in higher vertical orbital frequencies $\Omega_z$ for stars with small maximal orbital height $z_{\max}$. This in turn affects various dynamical effects, such as the presence or absence of orbital resonances (between vertical and radial motion) and the degree of phase-mixing in the vertical phase space. Thus, the modelling of phenomena related to the vertical structure of galactic discs, such as $z$-$v_z$ phase spirals and breathing or bending waves, are likely to be affected by the assumed vertical profile.

\lastpagefootnotes

Another result of this study are the cored-exponential profiles, for which we also give analytic mass models constructed by modifying the \cite{Kuzmin1956} disc. These models share with the exponential and sech$^2$ profiles the exponential decay with scale height $h$ at large $|z|$, but differ at small $|z|$, where they possess a core of near-constant density with adjustable width $w\le h$ (the sech$^2$ profile has core width $w=2h$). As the precise vertical profiles of galactic discs near $z=0$ are difficult to assess observationally and no theoretical foundations for either the exponential or sech$^2$ (or any other profile) exist\footnote{Of course, \cite{Spitzer1942} obtained the sech$^2$ profile for an isothermal distribution under the assumption of one-dimensional dynamics and gravity. But neither are these assumption very good \citep{Sarkar2020}, nor galactic discs expected to be isothermal.} (yet), these cored-exponential profiles are a useful addition to the dynamicist's tool box and allow to study the effect of a density core of any width $w\le h$.

Like the \citetalias{MiyamotoNagai1975} models, our new models are suitable bases for the construction of non-axisymmetric bar-shaped models by convolving them with a function $f(x)$ \citep{LongMurali1992} and in fact, we have added them to the suite of mass models provided by the \href{https://github.com/WalterDehnen/discBar}{\texttt{discBar}} code \citep{Dehnen2023}. 

One drawback of the \citetalias{MiyamotoNagai1975} models as well as our new models is the lack of realism of the radial profile, which closely follows that of the \citeauthor{Kuzmin1956} disc. While this profile resembles an exponential (with scale length $h_R=2a/3$) over some radial range, a purely exponential profile would be desirable. \cite{Smith2015} have shown that three \citetalias{MiyamotoNagai1975} models can be combined to have a radially near-exponential surface density profile over four scale lengths. Of course, this approach is also available to our new models
.


\bibliographystyle{mnras}
\bibliography{discPot}

\begin{thebibliography}{}
\makeatletter
\relax
\def\mn@urlcharsother{\let\do\@makeother \do\$\do\&\do\#\do\^\do\_\do\%\do\~}
\def\mn@doi{\begingroup\mn@urlcharsother \@ifnextchar [ {\mn@doi@} {\mn@doi@[]}}
\def\mn@doi@[#1]#2{\def\@tempa{#1}\ifx\@tempa\@empty \href {http://dx.doi.org/#2} {doi:#2}\else \href {http://dx.doi.org/#2} {#1}\fi \endgroup}
\def\mn@eprint#1#2{\mn@eprint@#1:#2::\@nil}
\def\mn@eprint@arXiv#1{\href {http://arxiv.org/abs/#1} {{\tt arXiv:#1}}}
\def\mn@eprint@dblp#1{\href {http://dblp.uni-trier.de/rec/bibtex/#1.xml} {dblp:#1}}
\def\mn@eprint@#1:#2:#3:#4\@nil{\def\@tempa {#1}\def\@tempb {#2}\def\@tempc {#3}\ifx \@tempc \@empty \let \@tempc \@tempb \let \@tempb \@tempa \fi \ifx \@tempb \@empty \def\@tempb {arXiv}\fi \@ifundefined {mn@eprint@\@tempb}{\@tempb:\@tempc}{\expandafter \expandafter \csname mn@eprint@\@tempb\endcsname \expandafter{\@tempc}}}

\bibitem[\protect\citeauthoryear{{Antoja} et~al.,}{{Antoja} et~al.}{2018}]{Antoja2018}
{Antoja} T.,  et~al., 2018, \mn@doi [\nat] {10.1038/s41586-018-0510-7}, \href {https://ui.adsabs.harvard.edu/abs/2018Natur.561..360A} {561, 360}

\bibitem[\protect\citeauthoryear{{Binney} \& {Sch{\"o}nrich}}{{Binney} \& {Sch{\"o}nrich}}{2018}]{BinneySchoenrich2018}
{Binney} J.,  {Sch{\"o}nrich} R.,  2018, \mn@doi [\mnras] {10.1093/mnras/sty2378}, \href {https://ui.adsabs.harvard.edu/abs/2018MNRAS.481.1501B} {481, 1501}

\bibitem[\protect\citeauthoryear{{Binney} \& {Tremaine}}{{Binney} \& {Tremaine}}{2008}]{BinneyTremaine2008}
{Binney} J.~J.,  {Tremaine} S.,  2008, {Galactic dynamics. 2nd ed}.
Princeton, NJ, Princeton University Press

\bibitem[\protect\citeauthoryear{{Ciotti}}{{Ciotti}}{2023}]{Ciotti2023}
{Ciotti} L.,  2023, \mn@doi [\mnras] {10.1093/mnras/stad2511}, \href {https://ui.adsabs.harvard.edu/abs/2023MNRAS.525.2758C} {525, 2758}

\bibitem[\protect\citeauthoryear{{Ciotti} \& {Pellegrini}}{{Ciotti} \& {Pellegrini}}{1996}]{Ciotti1996}
{Ciotti} L.,  {Pellegrini} S.,  1996, \mn@doi [\mnras] {10.1093/mnras/279.1.240}, \href {https://ui.adsabs.harvard.edu/abs/1996MNRAS.279..240C} {279, 240}

\bibitem[\protect\citeauthoryear{{Dehnen} \& {Aly}}{{Dehnen} \& {Aly}}{2023}]{Dehnen2023}
{Dehnen} W.,  {Aly} H.,  2023, \mn@doi [\mnras] {10.1093/mnras/stac3124}, \href {https://ui.adsabs.harvard.edu/abs/2023MNRAS.518.2651D} {518, 2651}

\bibitem[\protect\citeauthoryear{{Dehnen} \& {Binney}}{{Dehnen} \& {Binney}}{1998}]{Dehnen1998}
{Dehnen} W.,  {Binney} J.,  1998, \mn@doi [\mnras] {10.1046/j.1365-8711.1998.01282.x10.1111/j.1365-8711.1998.01282.x}, \href {https://ui.adsabs.harvard.edu/abs/1998MNRAS.294..429D} {294, 429}

\bibitem[\protect\citeauthoryear{{Dehnen} \& {Hasanuddin}}{{Dehnen} \& {Hasanuddin}}{2018}]{Dehnen2018}
{Dehnen} W.,  {Hasanuddin} 2018, \mn@doi [\mnras] {10.1093/mnras/sty1726}, \href {https://ui.adsabs.harvard.edu/abs/2018MNRAS.479.4720D} {479, 4720}

\bibitem[\protect\citeauthoryear{{Dobbie} \& {Warren}}{{Dobbie} \& {Warren}}{2020}]{Dobbie2020}
{Dobbie} P.~S.,  {Warren} S.~J.,  2020, \mn@doi [\ojap] {10.21105/astro.2003.05757}, \href {https://ui.adsabs.harvard.edu/abs/2020OJAp....3E...5D} {3, 5}

\bibitem[\protect\citeauthoryear{{Evans} \& {Bowden}}{{Evans} \& {Bowden}}{2014}]{EvansBowden2014}
{Evans} N.~W.,  {Bowden} A.,  2014, \mn@doi [\mnras] {10.1093/mnras/stu1113}, \href {https://ui.adsabs.harvard.edu/abs/2014MNRAS.443....2E} {443, 2}

\bibitem[\protect\citeauthoryear{{Fuchs} et~al.,}{{Fuchs} et~al.}{2009}]{Fuchs2009}
{Fuchs} B.,  et~al., 2009, \mn@doi [\aj] {10.1088/0004-6256/137/5/4149}, \href {https://ui.adsabs.harvard.edu/abs/2009AJ....137.4149F} {137, 4149}

\bibitem[\protect\citeauthoryear{{Juri{\'c}} et~al.,}{{Juri{\'c}} et~al.}{2008}]{Juric2008}
{Juri{\'c}} M.,  et~al., 2008, \mn@doi [\apj] {10.1086/523619}, \href {https://ui.adsabs.harvard.edu/abs/2008ApJ...673..864J} {673, 864}

\bibitem[\protect\citeauthoryear{{Kuijken} \& {Dubinski}}{{Kuijken} \& {Dubinski}}{1995}]{KuijkenDubinski1995}
{Kuijken} K.,  {Dubinski} J.,  1995, \mn@doi [\mnras] {10.1093/mnras/277.4.1341}, \href {https://ui.adsabs.harvard.edu/abs/1995MNRAS.277.1341K} {277, 1341}

\bibitem[\protect\citeauthoryear{{Kuzmin}}{{Kuzmin}}{1956}]{Kuzmin1956}
{Kuzmin} G.~G.,  1956, Astron. Zh., 33, 27

\bibitem[\protect\citeauthoryear{{Long} \& {Murali}}{{Long} \& {Murali}}{1992}]{LongMurali1992}
{Long} K.,  {Murali} C.,  1992, \mn@doi [\apj] {10.1086/171764}, \href {https://ui.adsabs.harvard.edu/abs/1992ApJ...397...44L} {397, 44}

\bibitem[\protect\citeauthoryear{{Miyamoto} \& {Nagai}}{{Miyamoto} \& {Nagai}}{1975}]{MiyamotoNagai1975}
{Miyamoto} M.,  {Nagai} R.,  1975, \pasj, \href {https://ui.adsabs.harvard.edu/abs/1975PASJ...27..533M} {27, 533}

\bibitem[\protect\citeauthoryear{{Mosenkov}, {Savchenko}, {Smirnov}  \& {Camps}}{{Mosenkov} et~al.}{2021}]{Mosenkov2021}
{Mosenkov} A.~V.,  {Savchenko} S.~S.,  {Smirnov} A.~A.,   {Camps} P.,  2021, \mn@doi [\mnras] {10.1093/mnras/stab2445}, \href {https://ui.adsabs.harvard.edu/abs/2021MNRAS.507.5246M} {507, 5246}

\bibitem[\protect\citeauthoryear{{Nagai} \& {Miyamoto}}{{Nagai} \& {Miyamoto}}{1976}]{NagaiMiyamoto1976}
{Nagai} R.,  {Miyamoto} M.,  1976, \pasj, \href {https://ui.adsabs.harvard.edu/abs/1976PASJ...28....1N} {28, 1}

\bibitem[\protect\citeauthoryear{{Plummer}}{{Plummer}}{1911}]{Plummer1911}
{Plummer} H.~C.,  1911, \mn@doi [\mnras] {10.1093/mnras/71.5.460}, \href {https://ui.adsabs.harvard.edu/abs/1911MNRAS..71..460P} {71, 460}

\bibitem[\protect\citeauthoryear{{Rojas-Ni{\~n}o}, {Read}, {Aguilar}  \& {Delorme}}{{Rojas-Ni{\~n}o} et~al.}{2016}]{RojasNino2016}
{Rojas-Ni{\~n}o} A.,  {Read} J.~I.,  {Aguilar} L.,   {Delorme} M.,  2016, \mn@doi [\mnras] {10.1093/mnras/stw846}, \href {https://ui.adsabs.harvard.edu/abs/2016MNRAS.459.3349R} {459, 3349}

\bibitem[\protect\citeauthoryear{{Sarkar} \& {Jog}}{{Sarkar} \& {Jog}}{2020}]{Sarkar2020}
{Sarkar} S.,  {Jog} C.~J.,  2020, \mn@doi [\mnras] {10.1093/mnras/staa2924}, \href {https://ui.adsabs.harvard.edu/abs/2020MNRAS.499.2523S} {499, 2523}

\bibitem[\protect\citeauthoryear{{Smith}, {Flynn}, {Candlish}, {Fellhauer}  \& {Gibson}}{{Smith} et~al.}{2015}]{Smith2015}
{Smith} R.,  {Flynn} C.,  {Candlish} G.~N.,  {Fellhauer} M.,   {Gibson} B.~K.,  2015, \mn@doi [\mnras] {10.1093/mnras/stv228}, \href {https://ui.adsabs.harvard.edu/abs/2015MNRAS.448.2934S} {448, 2934}

\bibitem[\protect\citeauthoryear{{Spitzer}}{{Spitzer}}{1942}]{Spitzer1942}
{Spitzer} L.,  1942, \mn@doi [\apj] {10.1086/144407}, \href {https://ui.adsabs.harvard.edu/abs/1942ApJ....95..329S} {95, 329}

\bibitem[\protect\citeauthoryear{{Toomre}}{{Toomre}}{1963}]{Toomre1963}
{Toomre} A.,  1963, \mn@doi [\apj] {10.1086/147653}, \href {https://ui.adsabs.harvard.edu/abs/1963ApJ...138..385T} {138, 385}

\makeatother
\end{thebibliography}

\begin{appendix}
\section{Surface density}
\label{app:surf}
Since $\Phi(R,z)=\Psi(R,\zeta)$ with $\nabla^2\Psi=0$, we have 
\begin{align}
    \label{eq:rho:Phi}
    4\pi G\rho &= \nabla^2\Phi =    
    \frac{\partial^2\Phi}{\partial z^2}-\frac{\partial^2\Psi}{\partial \zeta^2},
\end{align}
such that the face-on projections of the modified Kuzmin models have surface density
\begin{align}
    \Sigma(R) = \!\int_{-\infty}^\infty \!\rho\,\mathrm{d}z
    = - \frac1{2\pi G} \int_{0}^\infty \frac{\partial^2\Psi}{\partial \zeta^2}\,\mathrm{d}z
    = \frac{M}{2\pi } \int_{0}^\infty \left(\frac2{X^3}-\frac{3R^2}{X^5}\right)\,\mathrm{d}z,
\end{align}
since $\partial^2\Phi/\partial z^2$ integrates to zero. In general, this integral cannot be expressed in closed form, though for the \citetalias{MiyamotoNagai1975} model \cite{Ciotti2023} provides an expression involving elliptic integrals. Following \cite{Dehnen2023}, we compute this for any modifier $\zeta(z)$ numerically via Gauss Legendre quadrature after the substitution $t = z/\sqrt{z^2+R^2+s^2}$.

\section{velocity dispersion}
\label{app:sigma}
To solve the Jeans equation for self-gravitating modified Kuzmin models, we use equation~\eqref{eq:rho:Phi} to obtain
\begin{align}
    \rho\sigma_z^2 = \int_z^\infty \rho\, \frac{\partial\Phi}{\partial z} \,\mathrm{d} z 
    &= \frac1{4\pi G} \left[\int_z^\infty \frac{\partial^2\Phi}{\partial z^2} \frac{\partial\Phi}{\partial z}\,\mathrm{d}z - \int_\zeta^\infty \frac{\partial^2\Psi}{\partial\zeta^2} \frac{\partial\Psi}{\partial\zeta}\,\mathrm{d}\zeta \right]
    = 
    \frac1{8\pi G} \left[\int_z^\infty \!\! \frac{\partial}{\partial z}\! \left(\frac{\partial\Phi}{\partial z}\right)^{\!\!2} \mathrm{d}z -  \int_Z^\infty \!\! \frac{\partial}{\partial\zeta}\! \left(\frac{\partial\Psi}{\partial \zeta}\right)^{\!\!2} \mathrm{d}\zeta\right]
    \nonumber \\ &= 
    \frac1{8\pi G} \left[ \left(\frac{\partial\Psi}{\partial \zeta}\right)^{\!\!2} - \left(\frac{\partial\Phi}{\partial z}\right)^{\!\!2} \right] = \frac1{8\pi G} \left(\frac{\partial\Psi}{\partial\zeta}\right)^{\!\!2} (1-\zeta'^2) = \frac{GM^2}{8\pi} \frac{Z^2}{{X}^6}(1-\zeta'^2).
\end{align}
 
\section{A general recipe for modifiers}
\label{app:recipe}
In order to obtain a vertical profile which closely follows a given functional form $f(|z|)$, a modifier $\zeta(z)$ satisfying all our conditions is constructed as follows. First, we define the integral $F(z)\equiv \int_0^z f(t) \diff t$. Then
\begin{align}
    \zeta(z) &= \frac{F(\infty)}{f(0)} + \frac1{F(\infty)}\int_0^{|z|} F(t) \,\mathrm{d} t, &
    \zeta'(z) &= \sign(z) \frac{F(|z|)}{F(\infty)}, &
    \zeta''(z) &= \frac{f(|z|)}{F(\infty)}.
\end{align}
As an example, we apply this recipe to $f(z)=\mathrm{e}^{-z^2/2w^2²}$ for near-Gaussian vertical profiles, giving
\begin{align}
    \zeta_{\mathrm{G}} &= z \erf\frac{z}{\sqrt{2}\,w} + \sqrt{\frac2\pi} w\left[\mathrm{e}^{-z^2/2w^2²} + \frac\pi2 - 1\right], &
    \zeta_{\mathrm{G}}' &= \erf\frac{z}{\sqrt{2}w}, &
    \zeta_{\mathrm{G}}'' &= \sqrt{\frac2\pi} \frac1{w} \,\mathrm{e}^{-z^2/2w^2²}.
\end{align}
For this modifier, $\zeta_{\mathrm{G}}-\xi_{\mathrm{G}}\ge0$ everywhere, such that the resulting density is non-negative everywhere.

\end{appendix}
\end{document}